\documentstyle[floats,prd,aps]{revtex}

\def\NIM{Nucl.\ Instrum.\ \& Methods}
\def\NPB{Nucl.\ Phys.\ B}
\def\PLB{Phys.\ Lett.\ B}
\def\PRL{Phys.\ Rev.\ Lett.}
\def\PRD{Phys.\ Rev.\ D}
\def\ZPC{Z.\ Phys.\ C}
\def\etal{{\it et al.}}

\def\ups4s{$\Upsilon$(4S)}

\def\epluseminus{$e^+-e^-$}
\def\ckm{Cabibbo-Kobayashi-Maskawa }

\def\vub{$V_{ub}$}
\def\vubovcb{$|V_{ub}/V_{cb}|$}
\def\ar{\rightarrow}
\def\btou{$b\rightarrow u\ell\nu$}
\def\btoc{$b\rightarrow c\ell\nu$}
\def\btopilnu{$B\rightarrow\pi\ell\nu$}
\def\btorholnu{$B\rightarrow\rho\ell\nu$}
\def\btohlnu{$B\rightarrow h\ell\nu$}
\def\gampi{\Gamma(B^0\rightarrow\pi^-\ell\nu)}
\def\gamrho{\Gamma(B^0\rightarrow\rho^-\ell\nu)}

\input{psfig}

\begin{document}
\tighten
\draft

\title{Measurement of $V_{ub}$}

\author{M. A. Selen}

\address{
Department of Physics,
University of Illinois at Urbana-Champaign,\\
Urbana, Illinois 61801-3080, USA
}

\maketitle

\begin{abstract}
The most recent experimental measurements of the weak mixing angle
\vub~are summarized.  Inclusive and exclusive analysis techniques
are described, including a new exclusive analysis by the CLEO
collaboration measuring the $B\rightarrow \pi\ell\nu$ branching
ratio. Future prospects are discussed.
\end{abstract}

\section{Introduction}

The Standard Model of electroweak interactions is undeniably one of
the most successful theories in particle physics, accommodating
all observed phenomena to date. For all its glory,
the theory has the undesirable feature that most of its parameters,
although physically well motivated, must be determined empirically.

As an example, the way in which the weak interaction can mix quarks
of different flavor is described by the \ckm matrix\cite{cabibbo,km}.
The nine entries in this $3\times 3$ matrix contain information
about the relative strengths and phases with which the ($u,c,t$)
quarks couple to the ($d,s,b$) quarks, and their values
must be measured experimentally. Our present knowledge of the magnitudes
of these parameters are summarized below\cite{pdg}.

\begin{equation}
{\bf\rm V} =
\left(\matrix{
V_{ud} & V_{us} & V_{ub} \cr
V_{cd} & V_{cs} & V_{cb} \cr
V_{td} & V_{ts} & V_{tb} \cr}\right) =
\left(\matrix{
0.9747\ar 0.9759 & 0.218\ar 0.224 & 0.002\ar 0.005 \cr
0.218\ar 0.224 & 0.9738\ar 0.9752 & 0.032\ar 0.048 \cr
0.004\ar 0.015 & 0.030\ar 0.048 & 0.9988\ar 0.9995\cr}\right)
\label{ckmb}
\end{equation}

\noindent
Note that the extreme diagonal elements, $V_{ub}$ and $V_{td}$, are
the least well determined.
\medskip

In this paper the latest inclusive and exclusive measurements of
the ratio \vubovcb~will be discussed.  These were both made
by the CLEO collaboration, using data accumulated at the
Cornell Electron Storage Ring (CESR). In the CLEO-II detector, three
concentric tracking devices provide charged-particle momentum resolution of
$\sigma_p/p = 0.005+0.0015p$ ($p$ in GeV/c), and a 7800 crystal CsI
calorimeter provides neutral shower energy resolution of
$\sigma_E/E = 0.019 - 0.001E + 0.0035/E^{0.75}$.
More detailed information is available elsewhere\cite{cleodetector}.
\smallskip

The data sample used in the inclusive analysis consists of
$924~pb^{-1}$ accumulated with the CESR center of mass energy tuned to
the $\Upsilon$(4S) resonance, and $416~pb^{-1}$ accumulated at energies
below the $B\overline{B}$ production threshold.  The more recent
exclusive analysis was done using approximately a factor of two more data.

\section{Inclusive Measurements}

All determinations of \vub ~to date have been made by studying charmless
semileptonic decays of $B$ mesons produced in \epluseminus collisions
with center of mass energy at or near the $\Upsilon$(4S) resonance
\cite{cleo90,argus90,argus91,cleo93,jknelson}.
These are all measurements of the inclusive momentum-dependent rate of
leptons, $dN_\ell/dP_\ell$, from $B\rightarrow X_u\ell\nu$
decays\cite{leptons}.

Since the rate of \btou ~is very small, the main experimental
challenge in these analyses is the suppression of backgrounds.  For
inclusive \btou ~analyses, the three main sources of unwanted leptons are
\btoc~decays, other $B$ meson decays (for example
$B\rightarrow\psi X, \psi\rightarrow\ell^+\ell^-$), and continuum (non
$B\overline{B}$) events.  ``Fake'' leptons, for example $\pi$'s that
penetrate the detector iron and are misidentified as $\mu$'s, must also
be considered.

\subsection{\btoc~Suppression}

\begin{figure}[htb]
\vspace{0.in}
\begin{center}
\unitlength 1.0in
\begin{picture}(6.,2.5)(0,0)
\put(0.,0.)
{\psfig{bbllx=0in,bblly=0in,width=2.8in,file=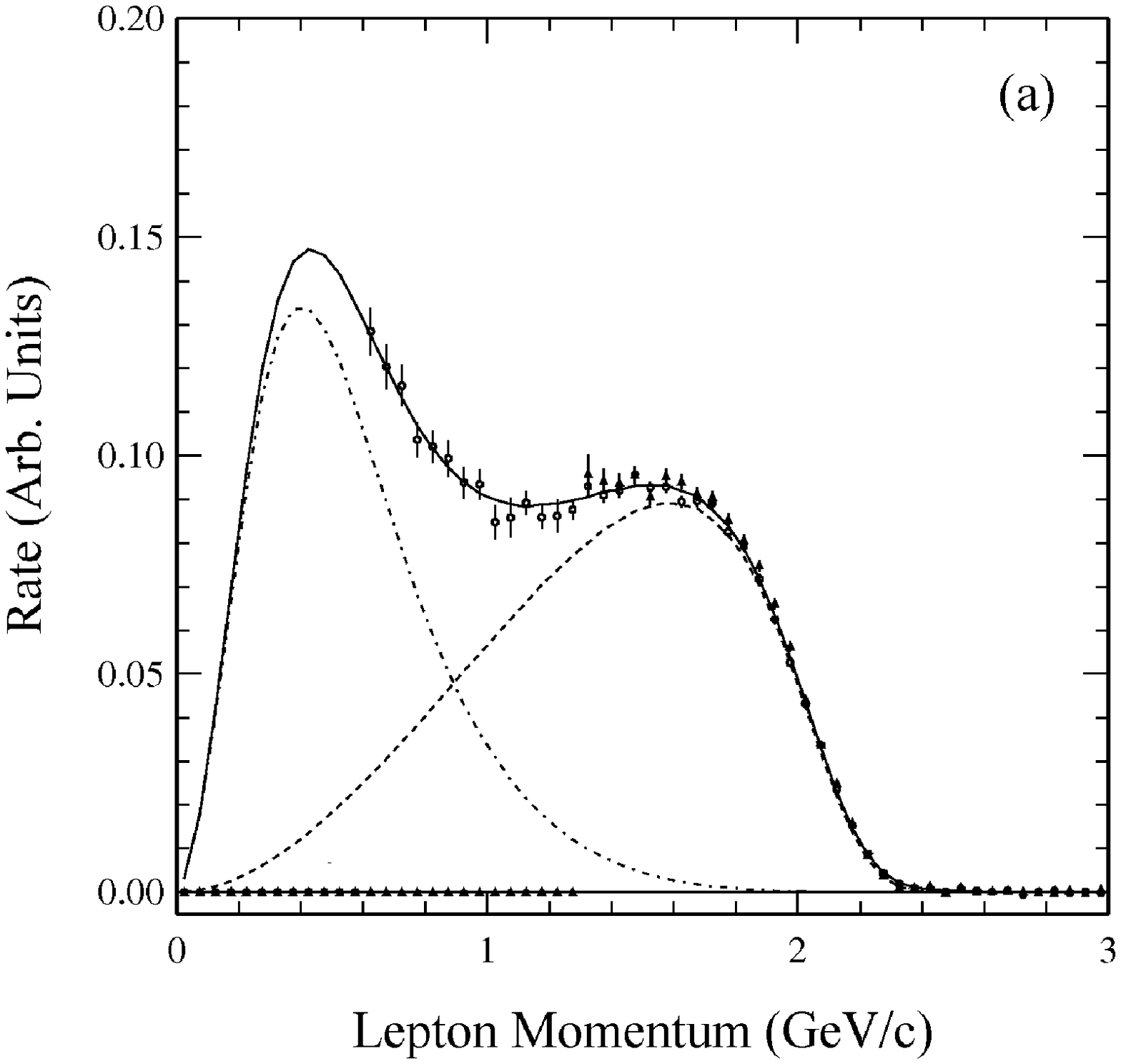}}
\put(3.,0.)
{\psfig{bbllx=0in,bblly=0in,width=2.8in,file=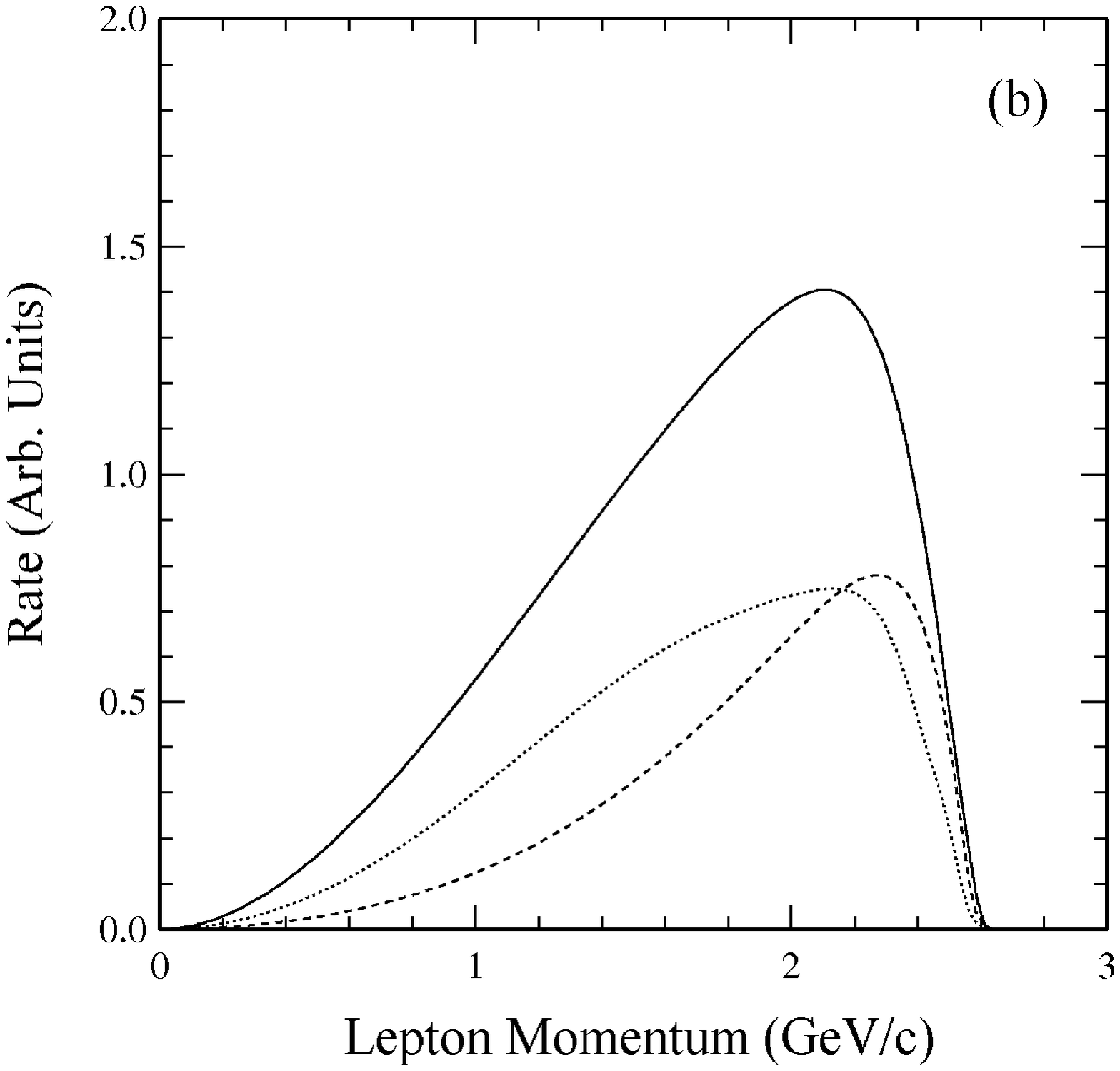}}
\end{picture}
\medskip
\caption{(a) Inclusive lepton momentum spectrum showing the
\btoc ~(dashed line) and $b\to c\to y\ell\nu$ (dot-dashed line) components,
as well as the total fit to CLEO data using the ACCMM model (solid line).
(b) The predicted \btou ~lepton momentum spectra for the ACCMM (solid line),
ISGW (dotted line) and WSB (dashed line) models.}
\label{leptonspectrum}
\end{center}
\end{figure}

The elimination of \btoc~decays from the \btou~sample is achieved
with a simple lepton momentum cut.  As can be seen in
Fig.~\ref{leptonspectrum}(a), the kinematic endpoint
momentum of leptons from $B\rightarrow X_c\ell\nu$ is
about 2.4 GeV/c\cite{batrest}. Fig.~\ref{leptonspectrum}(b) shows
theoretical lepton spectra for several models of
\btou\cite{accmm,isgw,wsb}.
Although the models differ significantly in detail,
they all share the basic kinematic feature that the lepton momentum
endpoint extends to $P_\ell\sim 2.6$ GeV/c.
\smallskip

The approach taken in all inclusive analyses has been to examine
the yield of leptons only in the endpoint region between
$\sim 2.4$ GeV/c and $2.6$ GeV/c\cite{lowerpcut}.  This restriction
eliminates most of the $b\rightarrow c\ell\nu$ contamination, but
only at the price of introducing a severe model-dependence in the
procedure used to extract \vub~from the data.  The reason for this
is clear upon examination of Fig.~\ref{leptonspectrum}(b).
The models shown differ considerably, both in overall
rate (plot area) and lepton momentum dependence (plot shape).
The effect of these differences are discussed in a later section.

\subsection{Continuum Suppression}

Leptons from non-$B\overline{B}$ continuum events, for example
$e^+e^-\rightarrow c\overline{c},~ c\rightarrow s\ell\nu$, are not
kinematically excluded from the inclusive \btou ~signal
region between 2.4 and 2.6 GeV/c.  This is a large source of background,
and is dealt with in two steps.
First, a set of continuum suppression requirements are used to eliminate
most of these events.  Second, the remaining background is removed by
subtracting the luminosity scaled lepton spectrum obtained by analyzing
``signal free'' data accumulated at center of mass energies below the
$B\overline{B}$ production threshold.
\smallskip

The most powerful continuum suppression requirements are topological
in nature, designed  to select the typically spherical $B\overline{B}$
events while rejecting the more ``jetty'' continuum background.
The shape variable used by CLEO is the Fox-Wolfram parameter
$R_2 = H_2/H_0$\cite{r2}.  The missing momentum of an event, $p_{miss}$,
provides additional discrimination against continuum processes.
This quantity should be large for $b\rightarrow u\ell\nu$ events where the
neutrino carries off appreciable momentum.  The requirements that $R_2 < 0.2$,
that $p_{miss} > 1~$GeV/c, and that the lepton and missing momentum point into
opposite hemispheres are used.  The net effect of these cuts is to
reduce the continuum background by a factor of 70 while maintaining
38\% efficiency for the $b\rightarrow u\ell\nu$ signal.
\smallskip

The statistical uncertainty introduced by the continuum subtraction is
a function of the size of the continuum data sample.  The normal operating
mode of CESR/CLEO-II is to accumulate data on the $\Upsilon$(4S) resonance
two-thirds of the time, and just below $B\overline{B}$ threshold
the remaining one-third.  The resulting continuum data sample is large enough
that the error introduced by the subtraction are much smaller than
the statistical error on the signal yield.

\subsection{Other Backgrounds}

After continuum suppression and subtraction, the events remaining in the
lepton momentum endpoint region are either true \btou ~signal or
non-continuum background.  This background is due to leptons from
$b\rightarrow c\rightarrow X\ell\nu$, from $B\rightarrow\psi$ and
$B\rightarrow\psi^\prime$ decays, and from fake leptons.  The fraction
of events in the lepton spectrum signal region due to these sources
is less than 10\%, is well understood from studying the data, and is
carefully accounted for when calculating the final \btou~yield.

\subsection{Signal Extraction and Model Dependence}

\begin{figure}[htb]
\vspace{0.in}
\begin{center}
\unitlength 1.0in
\begin{picture}(3.,3.)(0,0)
\put(-0.5,0.1)
{\psfig{bbllx=0pt,bblly=0pt,width=3.0in,height=3.0in,file=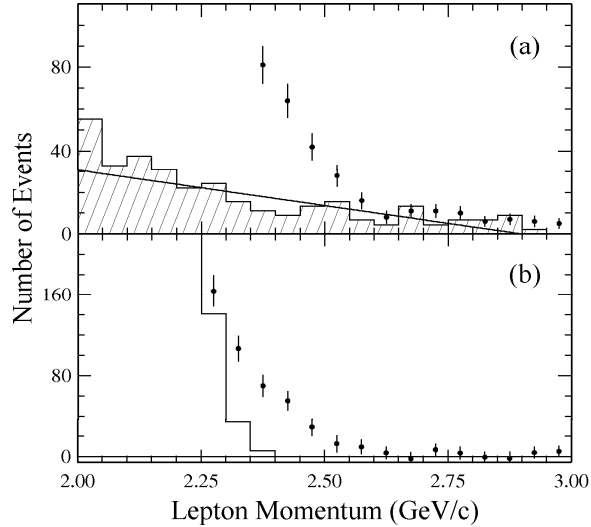}}
\end{picture}
\bigskip
\caption{CLEO-II inclusive lepton momentum distributions.
Shown in (a) are lepton spectra
from $\Upsilon$(4S) (filled circles) and continuum (hatched histogram) data,
as well as a fit to the continuum lepton distribution (solid line).
The filled circles in (b) show the result of subtracting the fitted and
scaled continuum distribution from the $\Upsilon$(4S) data.
The solid histogram shows the predicted contribution from \btoc ~decays.}
\label{endpoint}
\end{center}
\end{figure}

Fig.~\ref{endpoint}(a) shows lepton spectra from $\Upsilon$(4S)
and continuum data. The solid line is a fit to the continuum lepton
distribution.  Fig.~\ref{endpoint}(b) shows the result of luminosity
scaling and subtracting this fitted line from the $\Upsilon$(4S)
data, as well as the predicted contribution from \btoc ~decays.
A significant excess of events is seen, indicating the presence of
charmless B decays.

Extracting \vubovcb~ from the endpoint data involves several model
dependent factors:
\begin{equation}
\left|{V_{ub}\over V_{cb}}\right|={\Delta B_{ub}(p) \over d(p)B_{cb}}
\label{extvub}
\end{equation}
In this expression $B_{cb}$ is the \btoc ~branching
ratio and $\Delta B_{ub}(p)$ is the partial branching
ratio observed in the endpoint signal region\cite{pdependent}:
\begin{equation}
\Delta B_{ub}(p) = {N_{ub}(p)/\epsilon(p) \over 2N_{B\overline{B}}}
\label{partial}
\end{equation}
where $N_{ub}(p)$ is the number of \btou~events in the signal region,
$\epsilon(p)$ is the detection efficiency for these events,
and $2N_{B\overline{B}}$ is the number of $B$ mesons produced.
$\Delta B_{ub}(p)$ is somewhat model dependent since the detection efficiency
$\epsilon$ depends weakly on the shape of the lepton momentum spectrum.
\smallskip

The term $d(p)$ is the product of several strongly model-dependent
factors:
\begin{equation}
d(p) = f_u(p){\gamma_u \over \gamma_c}
\label{scalefact}
\end{equation}
where $f_u(p)$ is the fraction of the total \btou~spectrum in the
endpoint signal region (model dependence on spectrum shape),
and $\gamma_u$ relates the semileptonic
width and \vub~via $\Gamma_{b\rightarrow u\ell\nu} =
\gamma_u |V_{ub}|^2$\cite{gammac}, (model dependence on spectrum area).
\smallskip

Table \ref{incresults} shows the factors $d(p)$ and $\epsilon(p)$
as well the final extracted CLEO values of $\Delta B_{ub}(p)$
and $|V_{ub}/V_{cb}|$ for various models.  The model-dependence of
$|V_{ub}/V_{cb}|$ is larger than the experimental uncertainty,
and at present is the biggest factor limiting the accuracy of this
measurement.

\begin{table}[htb]
\caption{CLEO-II results for $d(p)$, $\epsilon(p)$, $\Delta B_{ub}(p)$
and $|V_{ub}/V_{cb}|$.  The values listed for $d(p)$ and $\epsilon(p)$
were calculated for the momentum range $2.4 < p_\ell({\rm GeV/c}) < 2.6$,
and are shown to illustrate the effect of model dependence.  The
values of $\Delta B_{ub}(p)$ and $|V_{ub}/V_{cb}|$ represent the
extended signal region $2.3 < p_\ell({\rm GeV/c}) < 2.6$.}
\label{incresults}
\begin{tabular}{lcccc}
Model & $d(p)$ & $\epsilon(p)$ & $10^6\times\Delta B_{ub}(p)$
& $|V_{ub}/V_{cb}|$ \\ \hline
ACCMM\cite{accmm}
& $0.12$ & $0.16\pm 0.01$ & $154\pm 22\pm 20$ & $0.076\pm 0.008$ \\
ISGW\cite{isgw}
& $0.05$ & $0.21\pm 0.02$ & $121\pm 17\pm 15$ & $0.101\pm 0.010$ \\
WSB\cite{wsb}
& $0.11$ & $0.20\pm 0.02$ & $122\pm 17\pm 16$ & $0.073\pm 0.007$ \\
KS\cite{ks}
& $0.19$ & $0.22\pm 0.02$ & $115\pm 16\pm 15$ & $0.056\pm 0.006$ \\
\end{tabular}
\end{table}

\section{Exclusive Measurements}

The CLEO collaboration has recently observed the exclusive charmless
semileptonic decay mode \btopilnu.  Measurement of this and other exclusive
channels will provide a new avenue for determining $|V_{ub}/V_{cb}|$ and
studying the \btou ~form factors, and should be a powerful tool for testing
the various available models.
\smallskip

This section will provide some details of the CLEO $B\rightarrow h\ell\nu$
analysis, where $h$ is $\pi^\pm$, $\pi^0$, $\rho^\pm$, $\rho^0$ or $\omega$.
Table \ref{exclpred} lists the theoretical predictions for the partial widths
of \btopilnu~ and \btorholnu, and their ratio.  Note again that the predicted
widths show a strong model-dependence.

\begin{table}[htb]
\caption{Predictions for the exclusive partial widths $\gampi$ and $\gamrho$.}
\label{exclpred}
\begin{tabular}{lccc}
Model                & $\gampi$~[$10^{12}|V_{ub}|^2$ sec$^{-1}$]
                     & $\gamrho$~[$10^{12}|V_{ub}|^2$ sec$^{-1}$]
                     & $\gamrho/\gampi$ \\ \hline
ISGW\cite{isgw}      & 2.1          &    8.3         & 4.0 \\
WSB\cite{wsb}        & 6.3 -- 10.0  &   18.7 -- 42.5 & 3.0 -- 4.3 \\
KS\cite{ks}          & 7.25         &   33.0         & 4.6 \\
ISGW II\cite{isgwii} & 9.6          &   14.2         & 1.5 \\
FGM\cite{faustov}    & $3.1\pm 0.6$ & $5.7\pm 1.2$   & $1.8\pm 0.5$ \\
\end{tabular}
\end{table}
\subsection{Neutrino Reconstruction}

The difficulty with exclusive reconstruction of semileptonic decays
is the lack of knowledge of the undetected neutrino's 4-momentum.
Using the large solid-angle coverage of the CLEO-II detector\cite{coverage}
to measure the {\it total} momentum and energy of the {\it rest} of
the event, CLEO is able to infer $(E_\nu,\vec{p}_\nu)$ of the neutrino
from the missing momentum and energy $(E_{miss},\vec{p}_{miss})$
of the event as a whole:
\begin{equation}
\label{eneutrino}
E_\nu\sim E_{miss} = 2E_{beam} - E^{tot}_{meas}~~{\rm and}~~
\vec{p}_\nu\sim \vec{p}_{miss} = - \vec{p}^{~tot}_{meas}
\end{equation}
where $E^{tot}_{meas}$ ($\vec{p}^{~tot}_{meas}$) is the total
measured energy (momentum) of all tracks and showers in the event.
Analysis of Monte Carlo generated \btopilnu~events yields
resolutions of 110 MeV and 260 MeV for $|\vec{p}_\nu|$ and $E_\nu$
respectively.
\smallskip

This method assumes the neutrino from \btou~is the only
undetected particle, making it crucial to reject events containing
additional unseen particles (neutrinos and/or $K_L$'s).  This is
accomplished by demanding that candidate events have only one
identified lepton, have zero net charge, and that the reconstructed
mass of the neutrino ($m_\nu^2=M^2_\nu - P^2_\nu$) be consistent with
zero\cite{zeromasscut}.
\smallskip

Additional constraints are placed on the final state hadrons.
Candidate $\pi^0$'s must have a
$2\gamma$ invariant mass within $2\sigma$ (about 12 MeV) of the nominal
$\pi^0$ mass, and $2\pi$ ($3\pi$) combinations must have invariant
mass within 90 MeV (30 MeV) of the $\rho$ ($\omega$) mass.
Identified leptons are required to have a momentum greater than
1.5 GeV/c (2.0 GeV/c) in the $\pi\ell\nu$ ($\rho/\omega\ell\nu$) modes.
\smallskip

For events passing the above neutrino, meson, and lepton selection
requirements, the reconstructed ``beam constrained'' $B$ mass
$m_B\equiv\sqrt{E^2_{beam}-|\vec{p}_h + \vec{p}_\ell + \vec{p}_\nu|^2}$
and energy difference $\Delta E\equiv E_{beam}-(E_h-E_\ell-|\vec{p}_\nu|)$
are calculated.  Real \btohlnu~events should have $\Delta E$ close to
zero and $m_B$ close to the $B$ meson rest mass.  Monte Carlo studies
are used to determine the optimum location of the signal region
in the $\Delta E-m_B$ plane, finding -250 $< \Delta E {\rm (MeV)} <$ 150
and 5.265 $< m_B {\rm (GeV/c^2)} <$ 5.2875.
\smallskip

As in the inclusive analysis, event shape variables\cite{r2} are used to
suppress continuum backgrounds, and a continuum subtraction is done.
The contribution from fake leptons is studied using data, and is
also subtracted.  The remaining background is predominantly due to
\btoc~decays containing an additional $\nu$ or $K_L$, and to cross-feed
from other \btou~modes.

%
\begin{figure}[htb]
\vspace{0.in}
\begin{center}
\unitlength 1.0in
\begin{picture}(3.,3.)(0,0)
\put(-2.0,0.0)
{\psfig{bbllx=0pt,bblly=0pt,height=2.5in,file=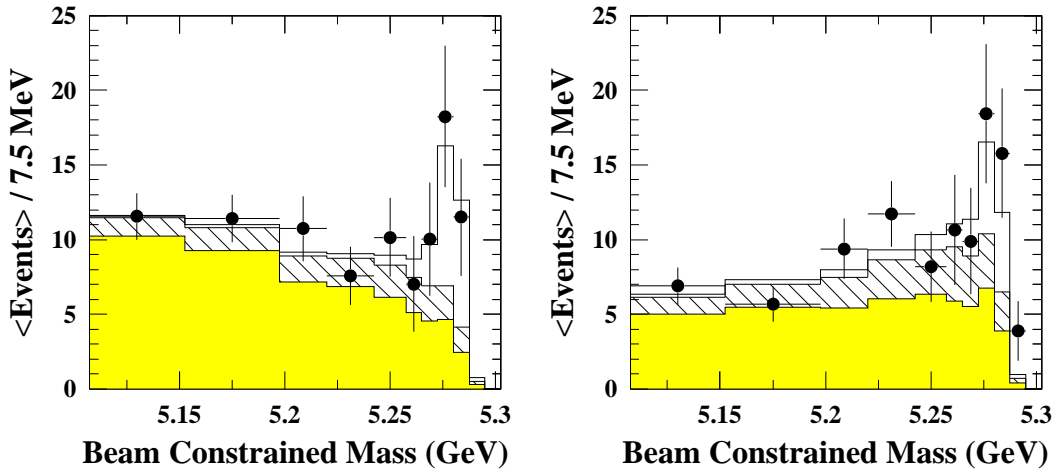}}
\end{picture}
\bigskip
\caption{Beam constrained mass distributions for combined $\pi\ell\nu$ (left)
and  $\rho\ell\nu$ (right) modes.  The points are continuum and fake
subtracted data, the histograms show the contribution from signal (hollow),
\btoc~(shaded), and \btou~cross-feed (hatched).}
\label{lawrence}
\end{center}
\end{figure}
%

\subsection{Signal Extraction and Model Dependence}

After subtracting the contributions from continuum and fake leptons and
selecting events in the $\Delta E$ signal region, the beam-constrained
mass ($m_B$) distributions of the five signal modes are simultaneously
fitted.  The shapes of the \btoc~and cross-feed background contributions
are obtained from Monte Carlo.  The normalization of the \btoc~background
for each mode are free parameters in the fit.  The isospin relations
${1\over 2}\Gamma(B^0\rightarrow\pi^-\ell^+\nu)=
\Gamma(B^+\rightarrow\pi^0\ell^+\nu)$ and
${1\over 2}\Gamma(B^0\rightarrow\rho^-\ell^+\nu)=
\Gamma(B^+\rightarrow\rho^0\ell^+\nu)\approx
\Gamma(B^+\rightarrow\omega\ell^+\nu)$ constrain the relative neutral
and charged meson rates.  The total signal and cross-feed background
for all five modes is in this way parameterized by two numbers $N_\pi$
and $N_\rho$, the total yield of $\pi$ events and $\rho$ events
respectively.
\smallskip

The beam-constrained mass distributions for \btopilnu ~and \btorholnu
{}~are shown in Fig.~\ref{lawrence}.  The results of the fit
for the case of \btopilnu~ are summarized in Table \ref{pifit}, where
the signal yield, efficiency, \btoc~ background, and \btou~ cross-feed
probabilities are shown for both the ISGW and WSB models.  Note that
although the yields are very similar for both models, the efficiencies
differ significantly.  The contribution from higher mass and non-resonant
\btou~modes, denoted ``other $u\ell\nu$'', is fixed by inclusive lepton
endpoint spectrum measurements.
\smallskip

Many checks have been performed to verify that the observed signal is real,
including fitting the energy difference ($\Delta E$) distribution rather
than the beam-constrained mass, examining the lepton momentum spectrum and the
distribution of angles between the pion and the lepton in the $W$ rest-frame.
More details of these studies, as well as a discussion of systematic errors,
can be found elsewhere\cite{moriond}.
\medskip

\begin{table}[htb]
\caption{Backgrounds, efficiencies and fit results for the \btopilnu~
analysis.}
\medskip
\label{pifit}
\begin{tabular}{lcccc}
& \multicolumn{2}{c}{$\pi^-\ell\nu$} & \multicolumn{2}{c}{$\pi^0\ell\nu$} \\
                 & ISGW         & WSB           & ISGW & WSB \\  \hline
Raw Data         & \multicolumn{2}{c}{30}          & \multicolumn{2}{c}{15} \\
Continuum Bkg.   & \multicolumn{2}{c}{$2.3\pm0.8$} &
\multicolumn{2}{c}{$1.0\pm0.5$} \\
Fake Lepton Bkg. & \multicolumn{2}{c}{$1.2\pm0.3$}   &
\multicolumn{2}{c}{$0.7\pm0.2$} \\
other $u\ell\nu$ Bkg. & \multicolumn{2}{c}{0.6} & \multicolumn{2}{c}{0.2} \\
Efficiency       & 2.9\%        & 2.1\%         & 1.9\%       & 1.4\% \\
Signal Yield     & $15.6\pm5.3$ & $16.3\pm5.3$  & $5.0\pm1.7$ & $5.3\pm1.7$ \\
$b\to c$ Bkg.    & $9.8\pm1.1$  & $9.8\pm1.1$   & $1.8\pm0.5$ & $1.7\pm0.5$ \\
$\rho/\omega$ Bkg.&$3.8\pm1.7$  & $3.4\pm1.4$   & $1.8\pm0.8$ & $1.6\pm0.7$ \\
\end{tabular}
\end{table}

Correcting the yield for efficiency results in a preliminary measurement of
the \btopilnu~ branching ratio: ${\cal B}(B^0\rightarrow\pi^-\ell\nu)=
(1.19\pm0.41)\times 10^{-4}$ using ISGW and $(1.70\pm0.55)\times 10^{-4}$
using WSB.  The errors shown are statistical only.
\smallskip

For the vector meson modes the fit results are used to calculate an
upper limit rather than a branching ratio since the non-resonant
contribution is uncertain.  Using the conservative assumption that there
is no non-resonant \btou~background present in the \btorholnu~signal
region, the 90\% confidence level upper limits for
${\cal B}(B^0\rightarrow\rho^-\ell\nu)$ is found using the ISGW (WSB)
model to be $<3.1\times 10^{-4}$ ($<4.6\times 10^{-4}$), consistent with
the upper limits previously published by CLEO\cite{cleoexclusive}.
\smallskip

One of the more interesting quantities that can be extracted from the data is
an upper limit on the {\it ratio} of branching ratios
${\cal B}(B^0\rightarrow\rho^-\ell\nu) / {\cal B}(B^0\rightarrow\pi^-\ell\nu)
< 3.4$ at the 90\% confidence level for both WSB and ISGW models.  This can be
directly compared to the ratio of partial widths found in Table \ref{exclpred}.

\section{Conclusions and Future Prospects}

At present, the limiting uncertainly in \vubovcb ~is theoretical.
The statistical error of the inclusive measurements is about 10\%, and the
variation between models is at least twice as large.  Even within a
single model, variation of parameters within reasonable limits can yield
significant changes\cite{hwang}.  CLEO is in the process of repeating its
inclusive analysis with over than a factor of two more data, which will
further decrease the statistical error of the endpoint analysis.
\smallskip

Experiments can do more, however, to provide guidance to the theoretical
community.  With sufficient statistics, measuring the $q^2$ distribution
of leptons in the inclusive endpoint region should provide useful
feedback.  The model dependence can in principle be reduced, or at the
very least explored, by examining data in different regions of the
$q^2-p_\ell$ plane.
\smallskip

Semileptonic $\Lambda_b$ decays may provide another avenue to study \vub.
A measurement of the form factors in $\Lambda_c\to \Lambda\ell\nu$ can be
related to $\Lambda_b\to p\ell\nu$ using SU(3) and HQET, and used to extract
\vubovcb\cite{datta}. Several authors have discussed ways of relating the
differential spectra for \btou~ and $b\to s\gamma$ to reduce the uncertainty
in the endpoint region due to hadronization\cite{falk,neubert,bigi,korchemsky}.
\smallskip

The statistical significance of exclusive measurements is still poor,
but will improve with experimental running.  It is unlikely that exclusive
analyses will ever surpass inclusive measurements in terms of raw statistical
accuracy, however they will certainly provide a very powerful tool for testing
various theoretical models.  The experimental limit on the ratio
${\cal B}(B^0\rightarrow\rho^-\ell\nu) / {\cal B}(B^0\rightarrow\pi^-\ell\nu)
< 3.4$ at 90\% confidence level is already slightly challenging for some.
Other measurements such as lepton momentum spectra, $q^2$ distribution,
and vector meson polarization will provide further theoretical tests
\cite{faustov2}.   Exclusive measurements at low $q^2$ may also prove
valuable\cite{akhoury}.  It has recently been shown that measurements
of \btorholnu ~and $B\to K^*\ell\overline{\ell}$ can be used to extract
\vub~ using SU(3) and HQET\cite{sanda}.
\smallskip

The theoretical problem of determining the form factors needed to calculate
\btou~is also being approached with lattice gauge calculations, and several
groups are making progress\cite{martinelli,ukqcd,ape,lanl}.
\medskip

In conclusion, good theoretical and experimental progress is being made
in the quest to determine \vubovcb, and the next few years should yield
significant advances in both.
\bigskip

I would like to thank Jeff Nelson, Ron Poling, Lawrence Gibbons and Ed
Thorndike for information about the latest CLEO analyses.  I would also
like to thank John Sloan and Aida El-Khadra for insight regarding lattice
calculations, and finally I would like to acknowledge Tom Browder and
Klaus Honscheid whose review paper ``B Mesons'' provided valuable
information and references.
\smallskip

\noindent I gratefully acknowledge the support of the Department
of Energy and the A. P. Sloan Foundation.


\begin{references}
\bibitem{leptons}
Both electrons and muons are used in these analyses.
\bibitem{batrest} This assumes that the $B$ meson is produced in the
reaction $e^+e^-\rightarrow\Upsilon{\rm(4S)}\rightarrow B\overline{B}$.
\bibitem{pdependent}
$(p)$ indicates that the quantity is a function of the endpoint
momentum interval chosen.
\bibitem{zeromasscut}
The requirement $M^2_\nu/2E_\nu < 350$ MeV is used since the
$M^2_\nu$ resolution varies approximately as $2E^2_\nu\sigma_{E_\nu}$.
\end{references}
\end{document}